\documentclass[conference]{IEEEtran}

\usepackage[english]{babel}
\usepackage{cite}
\usepackage{amsmath,amssymb,amsfonts,amsthm}
\usepackage{algorithm}
\usepackage{algorithmicx}
\usepackage{algpseudocode}
\usepackage{graphicx}
\usepackage{textcomp}
\usepackage{xcolor}
\usepackage{tikz}
\usepackage{url}
\usetikzlibrary{positioning}

\algblock{ParFor}{EndParFor}
\algnewcommand\algorithmicparfor{\textbf{parallel-for}}
\algnewcommand\algorithmicpardo{\textbf{do}}
\algnewcommand\algorithmicendparfor{\textbf{end\ parallel-for}}
\algrenewtext{ParFor}[1]{\algorithmicparfor\ #1\ \algorithmicpardo}
\algrenewtext{EndParFor}{\algorithmicendparfor}

\algnewcommand{\LineComment}[1]{\State \(\triangleright\) #1}
\AtBeginDocument{%
  \providecommand\BibTeX{{%
    \normalfont B\kern-0.5em{\scshape i\kern-0.25em b}\kern-0.8em\TeX}}}

\newtheorem{bmktheorem}{Theorem}[section]
\newtheorem{bmklemma}[bmktheorem]{Lemma}

\newcommand{\bigO}{\mathcal{O}}

\def\BibTeX{{\rm B\kern-.05em{\sc i\kern-.025em b}\kern-.08em
    T\kern-.1667em\lower.7ex\hbox{E}\kern-.125emX}}

\begin{document}

\title{Parallel, Portable Algorithms for Distance-2 Maximal Independent Set and Graph Coarsening}

\author{\IEEEauthorblockN{Brian Kelley\IEEEauthorrefmark{1} and
Sivasankaran Rajamanickam\IEEEauthorrefmark{2}}
\IEEEauthorblockA{Sandia National Laboratories, Albuquerque, New Mexico, U.S.A\\
Email: \IEEEauthorrefmark{1}bmkelle@sandia.gov,
\IEEEauthorrefmark{2}srajama@sandia.gov}}

\maketitle

\begin{abstract}

Given a graph, finding the distance-2 maximal independent set (MIS-2) of the vertices is a problem that is useful in several contexts such as algebraic multigrid coarsening or multilevel graph partitioning. Such multilevel methods rely on finding the independent vertices so they can be used as seeds for aggregation in a multilevel scheme. We present a parallel MIS-2 algorithm to improve performance on modern accelerator hardware. This algorithm is implemented using the Kokkos programming model to enable performance portability. We demonstrate the portability of the algorithm and the performance on a variety of architectures (x86/ARM CPUs and NVIDIA/AMD GPUs). The resulting algorithm is also deterministic, producing an identical result for a given input across all of these platforms. The new MIS-2 implementation outperforms implementations in state of the art libraries like CUSP and ViennaCL by 3-8x while producing similar quality results. We further demonstrate the benefits of this approach by developing parallel graph coarsening scheme for two different use cases. First, we develop an algebraic multigrid (AMG) aggregation scheme using parallel MIS-2 and demonstrate the benefits as opposed to previous approaches used in the MueLu multigrid package in Trilinos. We also describe an approach for implementing a parallel multicolor ``cluster'' Gauss-Seidel preconditioner using this MIS-2 coarsening, and demonstrate better performance with an efficient, parallel, multicolor Gauss-Seidel algorithm.

\end{abstract}

\maketitle

\section{Introduction}

Given an undirected graph $G = (V, E)$, an independent set is a subset $I_1$ of the vertices in $G$ such
that for any two vertices $u$, $v$ in $I_1$, the edge ($u$, $v$) is not in $E$. A \emph{maximal} independent set is an independent set where no additional vertex from $V$ can be added to $I_1$ while $I_1$ remains an independent set. Independent sets can be generalized to require an arbitrary distance of separation between any two vertices. Formally, a distance-$k$ independent set $I_k$ is subset of $V$, such that for any two vertices $u$, $v$ in $I_k$, there exists no path u$\leftrightarrow$v of length less than or equal to $k$ in $G$. We are interested in finding a \emph{maximal distance-2 independent set}, or MIS-2.

The problem of computing distance-2 maximal independent set can be used for graph coarsening in multilevel methods.
These are commonly employed in multigrid solvers \cite{muelu,Brandt86,XZ17},  domain decomposition methods \cite{FROSch},  graph or hypergraph partitioning \cite{boman2012zoltan,karypis1997parmetis}, and graph drawing \cite{ACE}.
Multilevel methods apply coarsening recursively until the graph is smaller than some threshold. Later, we describe two coarsening algorithms based on MIS-2.

As computer architectures are becoming increasingly diverse, typical computational science and engineering (CSE) applications are interested in supporting multiple CPU architectures (ARM, AMD, IBM, Intel) and multiple GPU architectures (AMD, Intel, NVIDIA). Such diversity has resulted in a focus on programming models such as Kokkos \cite{trott2021kokkos}. Kokkos provides hardware abstractions that make \emph{performance portable} software easier to develop. This refers to software that can utilize the extreme parallelism and hierarchical structure of GPUs, while still performing well on traditional multicore CPUs. We present such an algorithm for MIS-2 and implement it using Kokkos.

Non-deterministic parallel algorithms are common when designing parallel algorithms for modern accelerators because of several of the modern accelerator architectures do not guarantee deterministic behavior when using certain operations (e.g. atomics). However, the typical application user would like algorithms such as MIS-2 to be deterministic so that the behavior of solvers or graph partitioners can be replicated from run to run. Our parallel MIS-2 algorithm is deterministic across architectures and across several runs in the same architecture.

This MIS-2 algorithm can be used to develop a graph aggregation (coarsening) method. This graph aggregation method is also parallel and deterministic. We demonstrate its impact on two use cases related to linear solvers. The first use case is the multilevel aggregation in algebraic multigrid methods. The attributes of our algorithm (high parallelism and determinism) are both appealing for multigrid solvers. 
The focus in multigrid research has shifted towards faster setup times on accelerators in addition to solve times recently (either for better efficiency or because the problem structure changes and prevents reuse of setup). It has become important to have high-quality yet fast aggregation that can utilize modern accelerators. 
We will compare our approach against previous techniques to parallelize aggregation, like greedy distance-2 coloring, in section \ref{sec:aggregation-results}.

The second use case where we use the graph aggregation approach is in a Gauss-Seidel preconditioner. State of the art GPU implementations use coloring based approaches for parallel preconditioning. Coloring is primarily used to find independent rows in this use case. We use the new aggregation scheme to develop a Gauss-Seidel preconditioning algorithm that improves the runtime on all problems compared to past implementations \cite{MehmetColoring}.

The contributions of this paper can be summarized as:
\begin{itemize}
    \item We present a parallel MIS-2 algorithm that is deterministic and utilize large amounts of parallelism available in modern accelerator hardware.
    \item We demonstrate that the algorithm and the implementation is performance portable i.e. well suited for different accelerator hardware and standard CPUs. We show performance of the algorithm on two CPU architectures and two GPU architectures.
    \item Our implementation of this algorithm is 3-8x faster than implementations in libraries such as CUSP and ViennaCL, while producing outputs of similar quality.
    \item We develop a parallel aggregation algorithm using the MIS-2 algorithm as a kernel.
    \item We use the MIS-2 aggregation algorithm in multigrid setup and show this aggregation scheme enables faster solver convergence than the previous MIS2-based approach. The new approach still runs entirely on the accelerator and is also deterministic.
    \item Finally, we introduce cluster multicolor Gauss-Seidel, a parallel preconditioner that uses our parallel MIS-2 aggregation to improve convergence when compared to standard (point) multicolor Gauss-Seidel.
\end{itemize}

The rest of the paper is organized as follows.
Section \ref{sec:algo} describes our algorithm, as well as our aggregation scheme based on MIS-2. Section \ref{sec:clusterGS} introduces a new parallel Gauss-Seidel preconditioner that relies on graph coarsening (such as our MIS-2 aggregation). Section \ref{sec:theoretical} analyzes the asymptotic depth and work of the MIS-2 algorithm. Section \ref{sec:optimizations} describes the four algorithmic optimizations used to improve the performance of our algorithm. The results of our portable parallel algorithm and comparisons with other implementations are given in Section \ref{sec:results}.

\section{Related Work}
\label{sec:related}

MIS-2 has been previously applied to the problem of algebraic multigrid aggregation. In the ML multigrid package, Tuminaro and Tong \cite{ML} implemented MIS-2 using sparse matrix multiplication (SpGEMM) followed by a parallel MIS-1 algorithm \cite{Adams1998APM}. In terms of solve time and iterations, this aggregation scheme was found to perform comparably to their decoupled aggregation that runs sequentially within each processor's subdomain. Later, Bell, Dalton and Olson created a complete implementation of algebraic multigrid where every step of both setup and solve execute in parallel on an accelerator (specifically, an NVIDIA Tesla C2050 GPU) \cite{bell}. They present an efficient algorithm for computing an MIS-$k$ (for general $k \geq 1$) directly, without using SpGEMM. This algorithm was implemented in the CUSP library \cite{cusp}. For aggregation, Bell et al. use a simple coarsening strategy: each vertex in the MIS-2 is used as a root, and initial aggregates are formed from roots and their direct neighbors. Leftover vertices are joined arbitrarily to any adjacent aggregate. They note that for structured problems, this coarsening tends to produce irregularly shaped aggregates, increasing the number of solver iterations required. Both Bell's MIS-2 algorithm and this coarsening strategy were implemented in the ViennaCL library \cite{viennacl}. Additionally, Azad et al. presented a GraphBLAS-based formulation of these algorithms \cite{AzadMIS2}.

Outside of algebraic multigrid, Gilbert et al. evaluated MIS-2 coarsening for multilevel graph partitioning \cite{Gilbert}. Here, the MIS-2 based coarsening from Bell \textit{et al}. \cite{bell} is applied recursively to a graph until it is sufficiently small to use serial partitioning. 
Although heavy-edge matching (HEM), heavy-edge coarsening (HEC), and coarsening inspired by classical algebraic multigrid \cite{ClassicalMultigridCoarsening} are more commonly used in the literature for multilevel partitioning, Gilbert et al. found that MIS-2 coarsening outperforms HEM for regular graphs.

The related problem of MIS-1 has been focus of several studies. In particular, Luby designed two simple parallel algorithms for MIS-1. The first one (Monte Carlo Algorithm A \cite{luby}) is in fact the distance-1 analogue of our MIS-2 algorithm. We use this relationship to find the depth of our algorithm (Section \ref{sec:theoretical}). A later algorithm by Adams and Demmel achieved $\bigO(1)$ expected time for structured graphs in a PRAM model \cite{Adams1998APM}. Another problem is the \emph{maximum} independent set, or the largest \emph{maximal} independent set of a graph. This is a classic NP-hard problem, and both approximate and exact algorithms have been widely studied. Chang, Li and Zhang designed linear and near-linear approximate algorithms for maximum independent set using their Reducing-Peeling heuristic for iteratively reducing the size of the problem \cite{reducingPeeling}. Hespe, Schulz and Strash designed a parallel method for kernelization (exact reduction to a smaller problem) that can then be solved by exact or heuristic algorithms \cite{scalableKernelization}.

\section{Algorithms}
\label{sec:algo}
\subsection{MIS-2 Algorithm}


Algorithm \ref{alg:kk} is our parallel MIS-2 algorithm. It uses a similar idea as Bell's algorithm in the $k=2$ case \cite{bell}, but includes four important optimizations that greatly increase its performance. These will be discussed in detail in section \ref{sec:optimizations}.

\begin{algorithm}[t!]
\caption{MIS-2: Kokkos Kernels Algorithm (SIMD parallelism not shown)}
\label{alg:kk}
\begin{algorithmic}[1]
    \Procedure{MIS-2}{$G = (V,E)$}
        \LineComment{IN: vertex is in MIS}
        \State $\mathit{IN} \gets 0$
        \LineComment{OUT: vertex is not in MIS}
        \State $\mathit{OUT} \gets \mathit{UINT\_MAX}$
        \State $\mathit{worklist}_1 \gets 0\ldots|V|$
        \State $\mathit{worklist}_2 \gets 0\ldots|V|$
        \State $\mathit{iter} \gets 0$
        \While {$\mathit{worklist}_1 \neq \emptyset$}
            \LineComment{Refresh row status}
            \ParFor{$v \in \mathit{worklist}_1$}
                \LineComment{$|$ denotes bitwise concatenation}
                \LineComment{$h(...)$ is a hash function}
                \State $T_v \gets h(\mathit{iter}, v) | v+1$ 
            \EndParFor
            \LineComment{Refresh column status}
            \ParFor{$v \in \mathit{worklist}_2$}
                \State $M_v \gets min(T_w: w \in \mathit{adj}(v))$
                \If{$M_v = \mathit{IN}$}
                    \State $M_v \gets \mathit{OUT}$
                \EndIf
            \EndParFor
            \LineComment{Decide IN/OUT of set}
            \ParFor{$v \in \mathit{worklist}_1$}
                \If{$\exists w \in \mathit{adj}(v): M_w = \mathit{OUT}$}
                    \State $T_v \gets \mathit{OUT}$
                \EndIf
                \If{$\forall w \in \mathit{adj}(v): T_v = M_w$}
                    \State $T_v \gets \mathit{IN}$
                \EndIf
            \EndParFor
            \LineComment{Compact worklists with parallel prefix sums}
            \State $\mathit{worklist}_1 \gets \{v \in \mathit{worklist}_1: T_v \notin \{\mathit{IN}, \mathit{OUT}\}\}$
            \State $\mathit{worklist}_2 \gets \{v \in \mathit{worklist}_2: M_v \neq \mathit{OUT}\}$
            \State $\mathit{iter} \gets \mathit{iter} + 1$
        \EndWhile
    \State\Return{$\{v: T_v = \mathit{IN}\}$}
    \EndProcedure
\end{algorithmic}
\vspace{-0.1cm}
\end{algorithm}

In the algorithm, all vertices are initially undecided: they may later become either added to the MIS-2 (\emph{IN}) or not (\emph{OUT}). The algorithm iterates until no undecided vertices remain. First, priorities in the form of a 3-tuple $T_v = (\mathit{status}, \mathit{rand}, \mathit{ID})$ are assigned to each vertex. The possible values for status are $\emph{IN} < \emph{UNDECIDED} < \emph{OUT}$. $\mathit{rand}$ is a pseudo-random priority that is computed using a deterministic hash function $h$. The algorithm will work with any $h$, but the $h$ used in practice will be discussed in section \ref{sec:priorities}. Lastly, the $\emph{ID}$ of a vertex is simply its index in the graph. These tuples can be compared with each other in a lexicographic manner. First the $\mathit{status}$es are compared. If they are equal, the $\mathit{rand}$ values are compared. If they are also equal, the $\emph{ID}$ is compared last. Since each vertex ID is unique, this last comparison can never be a tie. To decide which vertices can be added to the set, the \emph{minimum} of $T_v$ within a radius-2 neighborhood of each vertex is computed. We use the same idea as Bell et al. to do this efficiently: if every vertex $v$ knows $M_v^k$ is the minimum tuple in a radius-$k$ neighborhood, then over all neighbors $w$, $v$ can compute $M_v^{k+1} = \mathit{min}(M_w^k)$. For MIS-2, this process can be repeated twice so that every vertex knows the minimum tuple in a radius-2 neighborhood. If $T_v = M_v^2$, $v$ has the lowest tuple in its neighborhood so it can be marked \emph{IN} without ambiguity. No other vertex in the neighborhood can also see that it has the minimum. Likewise, if $M_v^2 = (\mathit{IN}, *, *)$, $v$ is in the neighborhood of an \emph{IN} vertex so $v$ must be marked \emph{OUT}. Algorithm \ref{alg:kk} is designed to be as efficient as possible for MIS-2, so we only need to compute $M_v$ as the radius-1 minimum tuple, and then use $M_w$ among all neighbors $w$ to decide if $v$'s status can be decided. 
The main iteration has four phases: \emph{Refresh Row} (assign random priorities to vertices),   \emph{Refresh Column} (compute $M_v$ in distance-1 neighborhood), \emph{Decide Set} (compute status using distance-2 information) and maintaining the worklists for next iteration.
For more details, section \ref{sec:priorities} describes the hash function $h$ and how it was chosen, and section \ref{sec:compressedTuples} explains how tuples are stored in a compressed format for performance.

\begin{figure}
\begin{center}
\begin{tikzpicture}[
rednode/.style={circle, draw=red!90, align=center, fill=red!30, thick, minimum size=0mm, font=\scriptsize},
whitenode/.style={circle, draw=darkgray, align=center, fill=white, thick, minimum size=0mm, font=\scriptsize},
labelnode/.style={align=left, fill=white, minimum size=0mm, font=\normalsize},
]
\node[whitenode](n1){1 \\ 1, 1};
\node[whitenode](n2)[above=0.5cm of n1] {2 \\ 3, 2};
\node[whitenode](n3)[right=0.5cm of n2] {3 \\ 5, 3};
\node[whitenode](n4)[right=0.5cm of n3] {4 \\ 2, 4};
\node[whitenode](n5)[below=0.5cm of n4] {5 \\ 7, 5};
\node[whitenode](n6)[right=0.5cm of n4] {6 \\ 8, 6};
\node(label)[left=0.7cm of n2]{\emph{Refresh Row:}};
\draw[thick] (n1.north) -- (n2.south);
\draw[thick] (n2.east) -- (n3.west);
\draw[thick] (n3.east) -- (n4.west);
\draw[thick] (n4.east) -- (n6.west);
\draw[thick] (n5.north) -- (n4.south);
\end{tikzpicture}
\end{center}

\begin{center}
\begin{tikzpicture}[
rednode/.style={circle, draw=red!90, align=center, fill=red!30, thick, minimum size=0mm, font=\scriptsize},
whitenode/.style={circle, draw=darkgray, align=center, fill=white, thick, minimum size=0mm, font=\scriptsize},
labelnode/.style={align=left, fill=white, minimum size=0mm, font=\normalsize}
]
\node[whitenode](n1)              {1 \\ 1, 1};
\node[whitenode](n2)[above=0.5cm of n1] {2 \\ 1, 1};
\node[whitenode](n3)[right=0.5cm of n2] {3 \\ 2, 4};
\node[whitenode](n4)[right=0.5cm of n3] {4 \\ 2, 4};
\node[whitenode](n5)[below=0.5cm of n4] {5 \\ 2, 4};
\node[whitenode](n6)[right=0.5cm of n4] {6 \\ 2, 4};
\node(label)[left=0.7cm of n2]{\emph{Refresh Column:}};
\draw[thick] (n1.north) -- (n2.south);
\draw[thick] (n2.east) -- (n3.west);
\draw[thick] (n3.east) -- (n4.west);
\draw[thick] (n4.east) -- (n6.west);
\draw[thick] (n5.north) -- (n4.south);
\end{tikzpicture}
\end{center}

\begin{center}
\begin{tikzpicture}[
rednode/.style={circle, draw=red!90, align=center, fill=red!30, thick, minimum size=0mm, font=\scriptsize},
whitenode/.style={circle, draw=darkgray, align=center, fill=white, thick, minimum size=0mm, font=\scriptsize},
graynode/.style={circle, draw=darkgray, align=center, fill=gray, thick, minimum size=0mm, font=\scriptsize},
labelnode/.style={align=left, fill=white, minimum size=0mm, font=\normalsize}
]
\node[rednode](n1)              {1 \\ };
\node[whitenode](n2)[above=0.5cm of n1] {2 \\};
\node[whitenode](n3)[right=0.5cm of n2] {3 \\};
\node[rednode](n4)[right=0.5cm of n3] {4 \\ };
\node[whitenode](n5)[below=0.5cm of n4] {5 \\};
\node[whitenode](n6)[right=0.5cm of n4] {6 \\};
\node(label)[left=0.7cm of n2]{\emph{Decide Set:}};

\draw[thick] (n1.north) -- (n2.south);
\draw[thick] (n2.east) -- (n3.west);
\draw[thick] (n3.east) -- (n4.west);
\draw[thick] (n4.east) -- (n6.west);
\draw[thick] (n5.north) -- (n4.south);
\end{tikzpicture}
\end{center}

\begin{center}
\begin{tikzpicture}[
rednode/.style={circle, draw=red!90, align=center, fill=red!30, thick, minimum size=0mm, font=\scriptsize},
whitenode/.style={circle, draw=darkgray, align=center, fill=white, thick, minimum size=0mm, font=\scriptsize},
graynode/.style={circle, draw=darkgray, align=center, fill=gray, thick, minimum size=0mm, font=\scriptsize},
labelnode/.style={align=left, fill=white, minimum size=0mm, font=\normalsize}
]
\node[rednode](n1)              {1 \\ };
\node[whitenode](n2)[above=0.5cm of n1] {2 \\ 5, 2};
\node[whitenode](n3)[right=0.5cm of n2] {3 \\ 1, 3};
\node[rednode](n4)[right=0.5cm of n3] {4 \\ }; 
\node[whitenode](n5)[below=0.5cm of n4] {5 \\ 2, 5};
\node[whitenode](n6)[right=0.5cm of n4] {6 \\ 9, 6};
\node(label)[left=0.7cm of n2]{\emph{Refresh Row:}};

\draw[thick] (n1.north) -- (n2.south);
\draw[thick] (n2.east) -- (n3.west);
\draw[thick] (n3.east) -- (n4.west);
\draw[thick] (n4.east) -- (n6.west);
\draw[thick] (n5.north) -- (n4.south);
\end{tikzpicture}
\end{center}

\begin{center}
\begin{tikzpicture}[
rednode/.style={circle, draw=red!90, align=center, fill=red!30, thick, minimum size=0mm, font=\scriptsize},
whitenode/.style={circle, draw=darkgray, align=center, fill=white, thick, minimum size=0mm, font=\scriptsize},
graynode/.style={circle, draw=darkgray, align=center, fill=gray, thick, minimum size=0mm, font=\scriptsize},
labelnode/.style={align=left, fill=white, minimum size=0mm, font=\normalsize}
]
\node[rednode](n1)              {1 \\ }; 
\node[whitenode](n2)[above=0.5cm of n1] {2 \\ OUT};
\node[whitenode](n3)[right=0.5cm of n2] {3 \\ OUT};
\node[rednode](n4)[right=0.5cm of n3] {4 \\ };
\node[whitenode](n5)[below=0.5cm of n4] {5 \\ OUT}; 
\node[whitenode](n6)[right=0.5cm of n4] {6 \\ OUT}; 
\node(label)[left=0.7cm of n2]{\emph{Refresh Column:}};

\draw[thick] (n1.north) -- (n2.south);
\draw[thick] (n2.east) -- (n3.west);
\draw[thick] (n3.east) -- (n4.west);
\draw[thick] (n4.east) -- (n6.west);
\draw[thick] (n5.north) -- (n4.south);
\end{tikzpicture}
\end{center}

\begin{center}
\begin{tikzpicture}[
rednode/.style={circle, draw=red!90, align=center, fill=red!30, thick, minimum size=0mm, font=\scriptsize},
whitenode/.style={circle, draw=darkgray, align=center, fill=white, thick, minimum size=0mm, font=\scriptsize},
graynode/.style={circle, draw=darkgray, align=center, fill=lightgray, thick, minimum size=0mm, font=\scriptsize},
labelnode/.style={align=left, fill=white, minimum size=0mm, font=\normalsize}
]
\node[rednode](n1)              {1 \\};
\node[graynode](n2)[above=0.5cm of n1] {2 \\};
\node[graynode](n3)[right=0.5cm of n2] {3 \\};
\node[rednode](n4)[right=0.5cm of n3] {4 \\};
\node[graynode](n5)[below=0.5cm of n4] {5 \\};
\node[graynode](n6)[right=0.5cm of n4] {6 \\};
\node(label)[left=0.7cm of n2]{\emph{Decide Set:}};

\draw[thick] (n1.north) -- (n2.south);
\draw[thick] (n2.east) -- (n3.west);
\draw[thick] (n3.east) -- (n4.west);
\draw[thick] (n4.east) -- (n6.west);
\draw[thick] (n5.north) -- (n4.south);
\end{tikzpicture}
\end{center}
\caption{A full example of Algorithm \ref{alg:kk} computing an MIS-2 of a small graph. Red, gray and white nodes are \emph{IN}, \emph{OUT} and undecided, respectively. In each node, the first line is the vertex ID. After \emph{Refresh Row}, the second line contains the tuple $T_v = (\mathit{priority}, \mathit{ID})$. After \emph{Refresh Column}, the second line contains the tuple $M_v = (\mathit{priority}, \mathit{ID})$, or the minimum $T_v$ among all neighbors. Lastly, after \emph{Decide Set}, there are three possible cases for each vertex $v$. If for every neighbor $w$, $M_w = (*, v)$, then $v$ is marked \emph{IN}. If for any neighbor $w$, $M_w = \mathit{OUT}$, then $v$ is marked \emph{OUT}. Otherwise, $v$ remains undecided.}
\label{fig:example}
\vspace{-0.5cm}
\end{figure}
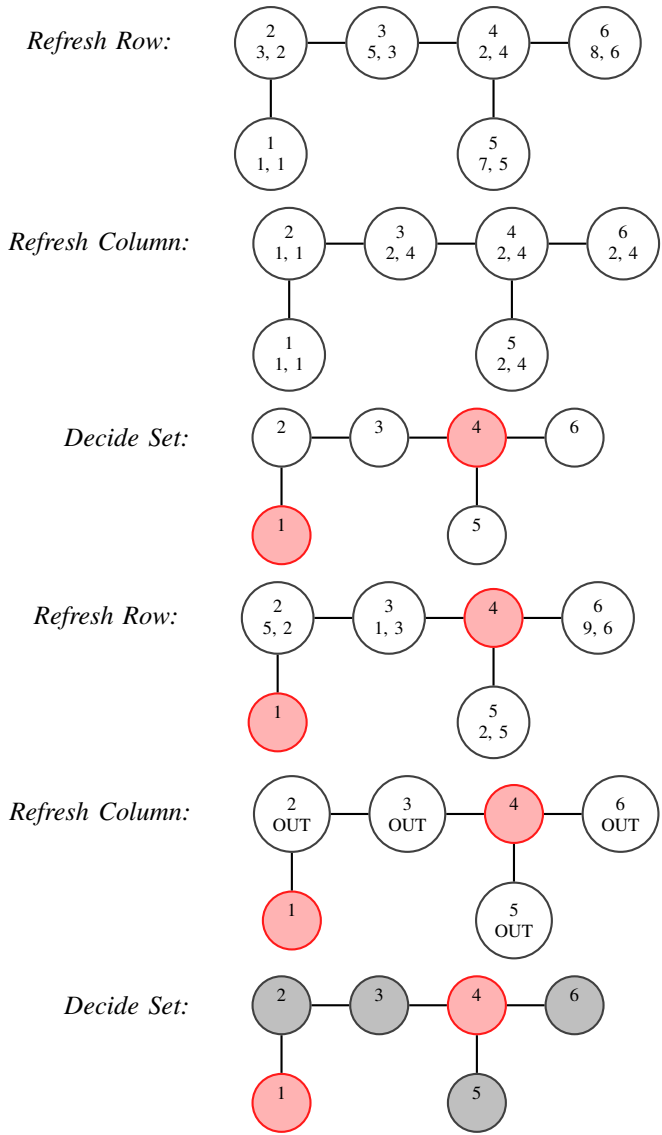

Fig. \ref{fig:example} is a complete visual example of Algorithm \ref{alg:kk} running on a small graph. Red nodes are in the MIS-2, and grey nodes are definitely not in the MIS-2. White nodes are still undecided. Each node contains two pieces of information: its ID, and a tuple $(\mathit{status}, \mathit{priority}, \mathit{ID})$. The status (in, out, or undecided) is represented with color, but the other two elements are shown as integers. Snapshots of the graph are shown at three points in the algorithm, for each of two iterations. \emph{Refresh Row} refers to line 15 in the algorithm, after $T_v$ (with a new random priority) has been assigned to each vertex. \emph{Refresh Column} is line 22, after $M_v$ has been computed as the minimum $T_w$ among all neighbors. Lastly, at \emph{Decide Set} (line 31), vertices have been marked either \emph{IN} or \emph{OUT} where possible. After two iterations, all vertices have been decided so the algorithm terminates and returns the MIS-2 $\{1, 4\}$.

\subsection{MIS-2 Based Aggregation}
\label{sec:mis2agg}

Given an MIS-2, it is straightforward to construct a graph coarsening suitable for smoothed-aggregation based algebraic multigrid (SA-AMG) or for a clustering based preconditioner.

Algebraic multigrid is a powerful preconditioning technique for the solution of a sparse linear system $Ax=b$. The method relies on building a hierarchy of smaller (coarser) linear systems from $A$. The residuals are restricted from a larger system to a smaller system, and the system is solved directly on the coarsest level. Finally, this solution is interpolated back to the original system.

\begin{algorithm}[t!]
\caption{Basic MIS-2 Coarsening}
\label{alg:coarsen}
\begin{algorithmic}[1]
    \Procedure{Coarsen}{$G = (V,E)$}
    \State $M \gets \mathit{MIS2}(G)$
    \ParFor{$v \in M$}
        \State Build aggregate from $v$ and its neighbors
    \EndParFor
    \ParFor{$v \in V$}
        \If{$v$ not aggregated}
            \State Add $v$ to the aggregate of any neighbor
        \EndIf
    \EndParFor
    \EndProcedure
\end{algorithmic}
\end{algorithm}
 
A simple method for aggregation with MIS-2 is shown in Algorithm \ref{alg:coarsen}. This was also the scheme used by Bell et al. \cite{bell}. Each vertex in the MIS-2 is made a root. Each root and its immediate neighbors become an aggregate. Then remaining vertices join with an adjacent aggregate arbitrarily. By the maximal property of MIS-2, all vertices must be within two edges of a root, so this scheme assigns every vertex to an aggregate.

\begin{algorithm}[t!]
\caption{MIS-2 based Aggregation}
\label{alg:agg}
\begin{algorithmic}[1]
    \Procedure{Aggregation}{$G = (V,E)$}
    \LineComment{Phase 1: Form initial aggregates using MIS-2}
    \State $M_1 \gets \mathit{MIS2}(G)$
    \ParFor{$v \in M_1$}
        \State Build aggregate from $v$ and its neighbors
    \EndParFor
    \LineComment{Phase 2: Form secondary aggregates}
    \State $M_2 \gets \mathit{MIS2}(G \setminus \{v \in V:$ v is aggregated $\})$
    \ParFor{$v \in M_2$}
        \If{$v$ has $\geq 2$ unagg. neighbors}
            \State Agg. $v$ with unagg. neighbors
        \EndIf
    \EndParFor
    \LineComment{Phase 3: Cleanup}
    \LineComment{Save current aggregate labels as ``tentative''}
    \State $\mathit{tent} \gets \mathit{labels}$
    \State $\mathit{coupling}(a, v) \gets |\{u : (u,v) \in E \land \mathit{tent}_u = a\}|$
    \State $\mathit{aggsize}(a) \gets |\{v : \mathit{tent}_v = a\}|$
    \ParFor{unagg. $v$}
        \State Join $v$ to agg. with max $\mathit{coupling}$
        \State If a tie, choose the agg. with min $\mathit{aggSize}$
    \EndParFor
    \EndProcedure
\end{algorithmic}
\end{algorithm}

In Kokkos Kernels, we introduce another approach for coarsening. Algorithm \ref{alg:agg} is portable, parallel version MIS-2 aggregation based on ML's sequential MIS-2 aggregation \cite{ML}. In the first phase, we build the initial set of aggregates from MIS-2 vertices and their immediate neighbors. In the second phase, we construct another MIS-2 from the subgraph induced by the non-aggregated vertices. The vertices in this MIS-2 can become aggregate roots if they have at least 2 unaggregated neighbors. Otherwise, we consider the aggregate too small and likely to cause increased fill-in during smoothing step of multigrid solver. Finally, phase 3 joins all leftover vertices to the most strongly coupled adjacent aggregate. Coupling to an aggregate $a$ is defined as the number of neighbors in $a$. To maintain determinism, the coupling and aggregate sizes are computed using the ``tentative'' labels from the end of phase 2. These remain constant during phase 3. Since algorithm \ref{alg:agg} computes an MIS-2 twice, its performance depends heavily on the performance of the MIS-2 implementation.

Later in section \ref{sec:aggregation-results}, we show that this aggregation scheme does enable faster solver convergence than algorithm \ref{alg:coarsen}. This effectively replicates the sequential quality results of ML \cite{ML} while running efficiently on modern accelerators like the NVIDIA V100.

\subsection{Cluster Multicolor Gauss-Seidel}
\label{sec:clusterGS}

The Gauss-Seidel (GS) method is commonly used as a preconditioner for Krylov methods like conjugate gradient (CG) and generalized minimum residual (GMRES). Classical GS is not readily parallelizable as each update to $x_i$ depends on all previous updates ($x_j, j < i$). However, using graph coloring, it is possible to find sets of vertices which are independent and can be updated in parallel in the same manner as GS \cite{MehmetColoring}. Here, we call this method point multicolor GS.

Point multicolor GS discovers parallelism at the cost of an increased number of iterations. We design a preconditioner that reduces the number of iterations, while speeding up both setup and solve. Our preconditioner coarsens the graph (e.g. using Algorithm \ref{alg:agg}) and colors the coarsened graph to yield sets of independent \emph{clusters}. Algorithm \ref{alg:cgs} shows how we use this structure for an improved preconditioner, we call \emph{cluster multicolor GS}. Rows within each cluster are updated sequentially, so locally this method is equivalent to classical GS. Therefore, cluster multicolor GS results in a preconditioner with a number of iterations closer to sequential Gauss-Seidel, alleviating the primary issue faced by point multicolor GS\cite{MehmetColoring}.

A symmetric version of this method (``SGS'') can be achieved by looping over the colors twice: first forward and then backward. Additionally, for the symmetric cluster method, the order of row updates within each cluster are reversed during the backward loop. 

\begin{algorithm}[t!]
\caption{Cluster Multicolor Gauss-Seidel}
\label{alg:cgs}
\begin{algorithmic}[1]
    \Procedure{CMGS}{$A, x, b$}
    \LineComment{Setup (reusable as long as A's structure unchanged)}
    \State $A_c \gets \mathit{coarsen}(A)$ \Comment{e.g. Algorithm \ref{alg:coarsen} or \ref{alg:agg}}
    \State $\mathit{clusters}_v \gets$ vertex in $A_c$ assigned to $v$
    \State $\mathit{colorsets} \gets \mathit{color}(A_c)$
    \LineComment{Apply}
    \For{$\mathit{color} \in \mathit{ncolors}$}
        \ParFor{$c \in \mathit{colorsets}_\mathit{color}$}
            \For{$i \in \mathit{clusters}_c$}
                \State $r \gets b_i - A_{i} x^\top$ \Comment {Row $i$ of $A$}
                \State $x_i \gets r/A_{ii}$
            \EndFor
        \EndParFor
    \EndFor
    \EndProcedure
\end{algorithmic}
\end{algorithm}

\section{Theoretical Analysis}
\label{sec:theoretical}

To find the asymptotic behavior of Algorithm \ref{alg:kk}, it is helpful to reduce the MIS-2 problem to MIS-1. This will allow us to apply theoretical results about a related MIS-1 algorithm. We will use a property of adjacency matrix exponentiation:

\begin{bmklemma}
\label{lemma:graphsquared} If $G$ is the adjacency matrix representation of a graph,
$G^k_{ij} \neq 0 \Leftrightarrow$ a path of length $k$ exists between $i$ and $j$ in $G$. Furthermore, if $G$ contains all self-loops, then $G^k_{ij} \neq 0 \Leftrightarrow$ a path of length $\leq k$ exists between $i$ and $j$ in $G$.
\end{bmklemma}

\begin{bmklemma}
\label{lemma:d1d2equivalence}
If $G$ is adjacency matrix representation of a graph (including all self-loops) and $I$ = MIS-1$\mathit{(G^2)}$, then $I$ is also a valid MIS-2 of $G$.
\end{bmklemma}

\begin{proof}
Let $I$ = MIS-1$(G^2)$. We want to show that $I$ satisfies both requirements (independence and maximality) for MIS-2 on $G$.

Assume $I$ violates the distance-2 independence property in $G$. Then there exist $i,j \in I$ such that there is a path of length $\leq 2$ between $i$ and $j$ in $G$. $G^2_{ij} \neq 0$ by \ref{lemma:graphsquared}. So $i,j$ are adjacent in $G^2$. Contradiction: $I$ is not an MIS-1 of $G^2$.

Now assume that $I$ is not a distance-2 maximal independent set of $G$. For some $v \notin I$, $I \cup \{v\}$ still does not violate distance-2 independence in $G$. Then $G$ contains no path of length $\leq 2$ between $v$ and any other vertex in $I$. For all $i \in I, G^2_{iv} = 0$ by \ref{lemma:graphsquared}. But this also means that $I \cup \{v\}$ satisfies distance-1 independence in $G^2$. Contradiction: $I \neq$ MIS-1$(G^2)$.

$I$ is an MIS-2 of $G$.
\end{proof}

By applying \ref{lemma:d1d2equivalence}, we can apply a theorem about Luby's Monte Carlo Algorithm A (from now on, simply called ``Luby's algorithm'') for MIS-1 \cite{luby} to our MIS-2 Algorithm \ref{alg:kk}. Luby's algorithm is the distance-1 analogue of algorithm \ref{alg:kk}. Initially, all vertices are candidates for the MIS-1 ($I = V$). Until $I$ converges, new random priorities are assigned to each vertex in $I$, and vertices with a higher priority neighbor are removed from $I$. The fact that vertices are not explicitly labeled \emph{IN} is a superficial difference: if $v$ has the highest priority among its neighbors, Luby's algorithm will remove $v$'s neighbors from $I$ which guarantees that $v$ can never be removed in subsequent iterations. $I$ therefore contains two mutually exclusive types of vertices: those which are undecided (analogous to $\mathit{worklist}_1$ in algorithm \ref{alg:kk}) and those marked \emph{IN}.
Note that in algorithm \ref{alg:kk}, we specify that a hash function $h$ is used as a pseudo-random number generator. The same function could be used in Luby's algorithm.
\emph{Assuming the same priorities are chosen for each vertex and iteration, Lemma \ref{lemma:d1d2equivalence} shows that Luby's algorithm run on $G^2$ will terminate in the same number of iterations as Algorithm \ref{alg:kk} run on $G$.} Since $G$ and $G^2$ have the same number of vertices, by Luby's Theorem 1 \cite{luby}, Algorithm \ref{alg:kk} can be expected to terminate in $\bigO(\log(V))$ iterations.
The parallel prefix sums have $\bigO(\log(V))$ depth and all other statements have $\bigO(1)$ depth, so the overall expected depth of Algorithm \ref{alg:kk} is $\bigO(\log^2(V))$.
We assume the work of a parallel prefix sum is $\bigO(n\log(n))$. An upper bound of the work per iteration of Algorithm \ref{alg:kk} (lines 9-30) is when both worklists contain all $V$ vertices. The loop on line 9 does constant work per vertex, while the loops on lines 14 and 20 inspect each neighbor, for a total of $\bigO(V+E)$ work. Combining these, an upper bound for the expected total work of Algorithm \ref{alg:kk} is $\bigO(\log V(V + E + V \log V))$ = $\bigO(V \log V + E \log V + V \log^2V)$.

\section{Algorithmic Optimizations}
\label{sec:optimizations}
\subsection{Pseudo-random Priorities}
\label{sec:priorities}
Both our Algorithm \ref{alg:kk} and Bell's algorithm \cite{bell} assign pseudo-random priorities to each vertex to determine an order for vertices to join the MIS-2. The difference is that Bell et al. algorithm chooses priorities once and uses them in every iteration, but the Kokkos Kernels algorithm computes new priorities during each iteration. Luby's Monte Carlo Algorithm A for MIS-1 similarly used new priorities each iteration \cite{luby}.

When fixed priorities are used, vertices may become part of a dependency chain. For example, suppose $v$ has a distance-2 neighbor $w$, and in $v$'s radius-2 neighborhood, $w$ and $v$ have the lowest and second-lowest priorities respectively. $v$'s membership in the MIS-2 cannot be decided until after $w$ is decided. If $w$ is decided to be \emph{IN}, then $v$ immediately becomes OUT (and vice versa). Meanwhile, no other vertices in $v$'s neighborhood could become \emph{IN} (since they all have higher priority than $v$). Short dependency chains are inevitable, but long dependency chains can serialize the entire algorithm (allowing only a single vertex to be decided every two iterations).

\begin{table}[h]
\caption{MIS-2 iteration counts for three random priority methods. Fixed is same as \cite{bell}. Xor and Xor* are used with Algorithm \ref{alg:kk}.}
\begin{center}
\begin{tabular}{l|llll}
                 & Fixed & Xor Hash & Xor* Hash \\ \hline
af\_shell7       & 11             & 23        & 8    \\
ecology2         & 12             & 11        & 8    \\
Hook\_1498       & 14             & 26        & 11   \\
PFlow\_742       & 14             & 39        & 12   \\
thermal2         & 12             & 17        & 9    \\
apache2          & 13             & 21        & 10   \\
Elasticity3D\_60 & 13             & 23        & 10   \\
Fault\_639       & 13             & 26        & 10   \\
Laplace3D\_100   & 14             & 20        & 10   \\
Serena           & 14             & 22        & 11   \\
tmt\_sym         & 12             & 18        & 8    \\
audikw\_1        & 14             & 22        & 10   \\
Emilia\_923      & 13             & 20        & 11   \\
Geo\_1438        & 14             & 26        & 11   \\
ldoor            & 11             & 16        & 8    \\
parabolic\_fem   & 11             & 9         & 9    \\
StocF-1465       & 14             & 28        & 10  
\end{tabular}
\label{tab:rng-iterations}
\end{center}
\end{table}

Table \ref{tab:rng-iterations} shows the number of iterations required by three different schemes for choosing the priorities. The test graphs are described in Section \ref{sec:results}. The iteration count is the number of times the loop on line 8 of Algorithm \ref{alg:kk} is executed. ``Fixed'' is the same as Bell's algorithm, where priorities are randomized once. ``Xor Hash'' and ``Xor* Hash'' are both versions of Algorithm \ref{alg:kk}, where a deterministic hash function $h$ is used as the pseudo-random generator. In order to give completely different values in different iterations, $h$ combines the vertex ID and iteration number: $h(\mathit{iter}, v) = f(f(iter) \oplus f(v))$, where $\oplus$ is bitwise XOR. For ``Xor Hash'', $f(x)$ is 64-bit xorshift, and for Xor*, $f(x)$ is the 64-bit xorshift* (xorshift followed by a linear congruential step). Both of these functions were discovered by Marsaglia \cite{xorshift}.
Surprisingly, xorshift is a poor hash function for this algorithm. It usually leads to a significantly higher iteration count than ``Fixed'', implying that $f(f(iter) \oplus f(v))$ is correlated from one iteration to the next, preventing a dependency chain from being broken. 
However, xorshift* does not have this issue. We use it in the implementation of Algorithm \ref{alg:kk} and all experiments in the rest of the paper.

\subsection{Worklists} \label{sec:worklists}
Bell et al. \cite{bell} process all vertices in the graph in every iteration. This causes a large amount of redundant work to be performed on vertices whose MIS-2 status is already known, especially in the later iterations when only a few vertices remain undecided.
To avoid redundant work, Algorithm \ref{alg:kk} maintains two distinct worklists for vertices. $\mathit{worklist}_1$ simply contains the undecided vertices. $\mathit{worklist}_2$ contains the vertices which are not adjacent to any \emph{IN} vertices. This is because of the behavior of the loop on line 14: once any vertex $v$ gets an \emph{IN} neighbor, $M_v$ will permanently become \emph{OUT}.

In Algorithm \ref{alg:kk}, on lines 29 and 30, the vertices which will not be processed again are filtered out using a parallel prefix sum (also known as a ``scan''). Kokkos provides a parallel scan that is efficient on all supported backends \cite{trott2021kokkos}. Deveci et al. previously used the scan in Kokkos to maintain edge worklists for greedy graph coloring \cite{MehmetColoring}.

\subsection{Compressed Status Tuples} \label{sec:compressedTuples}
Bell's algorithm uses tuples containing 3 elements: a \emph{status}, a \emph{priority}, and an \emph{ID}. The algorithm must store three of these tuples per vertex ($S, T$ and $\hat{T}$). The status can only take one of three values: \emph{IN}, \emph{OUT} or \emph{UNDECIDED}. The priority field can be an integer of any width. If it is narrow (e.g. 8 bits) ties become likely, but the unique ID is also compared as a tiebreak. Finally, the ID is an integer which must be wide enough to represent $|V|$. A straightforward implementation could use a 3-element structure to represent each field individually, but this is not the most efficient way to represent the information contained in the tuple as a whole.

Algorithm \ref{alg:kk} compresses these tuples into a single integer with the same width as the vertex IDs (typically 32 bits). \emph{IN} and \emph{OUT} are represented as special values 0 and UINT\_MAX respectively. This is correct because it maintains the ordering $\mathit{IN} < \mathit{UNDECIDED} < \mathit{OUT}$. No particular ordering needs to exist between two \emph{IN} or two \emph{OUT} tuples.

For \emph{UNDECIDED} vertices, the priority and vertex ID are packed together to occupy the range between \emph{IN} and \emph{OUT}. $b = \lceil \log_2(|V| + 2) \rceil$ bits are used to represent the ID component, and the remaining bits (e.g. $32-b$) are used for the priority. The vertex ID still functions as a tiebreak, since compressed tuples with different IDs must differ in at least one bit. The final value is computed as \verb^(priority << b) | (id + 1)^. The smallest possible value produced by this formula is 1 (for priority and id both 0). The largest possible value is
\verb^(0xFFFF... << b) | (maxVert + 1)^.
Vertices use 0-based numbering, so $\mathit{maxVert} + 1 = |V|$.

\begin{equation}
\label{eq:max-tuple}
\begin{gathered}
b \geq \log_2(|V| + 2) \\
\Rightarrow 2^b \geq |V| + 2 \\
\Rightarrow 2^b > |V| + 1 \\
\Rightarrow 2^b - 1 > |V|
\end{gathered}
\end{equation}

Equation \ref{eq:max-tuple} shows that there must be at least one zero bit in the least significant $b$ bits. Therefore, no combination of priority and vertex ID can collide with either \emph{IN} or \emph{OUT}. The technique of packing several pieces of information into a single integer was inspired by ECL-MIS \cite{ecl-mis}, a
parallel greedy algorithm for computing the MIS-1. In this algorithm, Burtscher et al. represented vertex status using 8-bit integers. If the least significant bit is 0, the vertex has been decided and the remaining bits will all be 1 for \emph{IN} and 0 for \emph{OUT}. If the least significant bit is 1, the remaining bits represent the randomized priority (and must be neither all 0 nor all 1). One important difference between ECL-MIS and our MIS-2 algorithm is that ECL-MIS does not require the priorities to be unique. Since our MIS-2 does, we include the vertex ID in the least significant bits.

\subsection{SIMD Parallelism} \label{sec:simd}
The final optimization used in Kokkos Kernels MIS-2 is the use of SIMD parallelism in the innermost loops. This level of parallelism is called ``warp-level'' or ``wavefront-level'' in NVIDIA and AMD terminology, respectively. It is not explicitly shown in Algorithm \ref{alg:kk}, but it is used in each loop over vertex neighbors (lines 15, 21 \& 24). The Kokkos SIMD-level parallel reduce functionality is used to compute the functions $\mathit{min}$, $\forall$ and $\exists$ over the neighbors of a vertex \cite{trott2021kokkos}.

The primary performance benefit of iterating on neighbors with SIMD-level is in the memory access pattern on GPUs. In the CRS sparse matrix/graph format, the adjacency list of each vertex is stored contiguously. When reading from this list with warp- or wavefront- level parallelism, memory accesses can be coalesced into a single memory transaction. This principle has been exploited previously in other graph problems, like breadth-first search \cite{MerrillSIMD} and betweenness centrality \cite{BetweennessCentralitySIMD}. However, this does not give a significant speedup on CPUs because of cache benefits. Additionally, the adjacency lists have irregular sizes and are not aligned to the SIMD register width (e.g. 256 bits for AVX), so most CPU SIMD instructions could not operate at full efficiency.

Even on GPUs, SIMD parallelism incurs some overhead. Every thread in the warp/wavefront must read some common information about the row, and during this time all the threads are busy with no useful parallelism. It is faster to iterate over its neighbors sequentially with a single GPU thread for low-degree vertices. In practice, \emph{we use SIMD parallelism only on GPUs and only if the average vertex degree is at least 16}.

\subsection{Lessons for portable, parallel graph algorithms}

The four optimizations described here are not only useful for MIS-2, but could be applied to a wide variety of parallel graph algorithms.

In Section \ref{sec:priorities}, we showed that for our pseudo-randomized algorithm, the choice of hash function can drastically affect the number of iterations of MIS-2. It was not obvious that xorshift was suboptimal until xorshift* was tested. For any algorithm that relies a hash being statistically independent with respect to the input, it is worthwhile to test different functions and verify that they are in fact independent.

The benefits of using worklists as in section \ref{sec:worklists} are clear for our MIS-2 algorithm, but in general this technique comes with tradeoffs. The parallel scan used to compact each worklist between iterations costs time, and reading an item (e.g. vertex ID) out of a worklist delays any useful work on that item. This approach has been applied successfully to greedy coloring \cite{MehmetColoring}, but it could also be applied to iterative refinement where only a subset of vertices (i.e. those on a boundary) are being considered.

The use of compressed status tuples discussed in section \ref{sec:compressedTuples} can be generalized as the elimination of redundant information. In our algorithm, a vertex's state may be decided (\emph{IN} or \emph{OUT}), or undecided. Only if it is undecided does it also need the random priority and ID. This \emph{conditional need for information} is exploited in the way we represent the vertex state tuples.

Finally, the use of SIMD parallelism is a key pattern for achieving high performance on GPUs. A very common pattern in graph algorithms is to iterate over a vertex's neighbors. If there are no data dependencies between the neighbors (as in our algorithm), SIMD increases both parallelism and improves memory coalescing as described in \ref{sec:simd}.

\section{Experimental Results}
\label{sec:results}

In this section, we evaluate the performance of our implementation of the MIS-2 algorithm in Kokkos Kernels. We begin by showing the impact of each one of our optimizations compared to the baseline implementation on V100 GPUs. The baseline is our implementation of the Bell et al. \cite{bell} algorithm using Kokkos. Next we demonstrate the scaling of the algorithm on mesh-like problems in terms of the MIS-2 size and the number of iterations. We then demonstrate portability betwen the NVIDIA V100 and AMD Radeon MI100 GPUs, and scalability on Intel and ARM CPUs. We also compare our implementation to the CUSP and ViennaCL libraries on V100 GPUs and demonstrate that our implementation achieves similar quality but better performance. We demonstrate the efficacy of MIS-2 based aggregation Algorithm \ref{alg:agg} by using it in the multigrid solver MueLu. Lastly, we show the performance of the cluster Gauss-Seidel preconditioner, Algorithm \ref{alg:cgs}. 
All experiments use the same set of 17 matrices. Two are generated using the Galeri and MueLu packages of Trilinos \cite{muelu}: Laplace3D\_100 is a $100^3$ grid with a 7-point stencil, and Elasticity3D\_60 is a $60^3$ grid with a 27-point stencil and 3 degrees of freedom per point. The other 15 matrices are from the Suite Sparse collection \cite{SuiteSparse}.

\begin{table*}[ht]
\caption{Summary statistics of matrices used in experiments, and mean times to run Algorithm \ref{alg:kk} on them for the four architectures. Times are in milliseconds and were averaged over 100 trials. Intel Skylake and ThunderX2 ARM use 48 and 56 OpenMP threads, respectively.}
\centering
\begin{tabular}{l|cccc|cccc}
                 & $|V| (\times 10^6)$     & $|E| (\times 10^6)$  & Avg deg. & Max deg. & NVIDIA V100 & AMD MI100 & Intel Skylake & ThunderX2 ARM \\ \hline
af\_shell7       & 0.505 & 9.047  & 17.92      & 35   & 3.55 & 4.75 & 4.90 & 6.47 \\
apache2          & 0.715 & 2.767  & 3.87       & 4    & 2.71 & 3.44 & 4.37 & 4.73 \\
audikw\_1        & 0.944 & 39.298 & 41.64      & 114  & 8.42 & 16.3 & 49.6 & 57.7 \\
ecology2         & 1.000 & 2.998  & 3          & 3    & 2.95 & 3.05 & 4.84 & 5.09 \\
Elasticity3D\_60 & 0.648 & 50.758 & 78.33      & 81   & 5.90 & 11.3 & 14.3 & 20.2 \\
Emilia\_923      & 0.923 & 20.964 & 22.71      & 48   & 6.84 & 9.44 & 18.7 & 17.8 \\
Fault\_639       & 0.639 & 14.627 & 22.9       & 114  & 5.07 & 7.05 & 9.18 & 13.3 \\
Geo\_1438        & 1.438 & 32.297 & 22.46      & 48   & 9.95 & 13.2 & 32.0 & 27.9 \\
Hook\_1498       & 1.498 & 31.208 & 20.83      & 57   & 10.1 & 13.9 & 19.0 & 29.5 \\
Laplace3D\_100   & 1     & 6.94   & 6.94       & 7    & 3.34 & 4.21 & 6.21 & 6.71 \\
ldoor            & 0.952 & 23.737 & 24.93      & 49   & 6.18 & 11.7 & 19.2 & 18.8 \\
parabolic\_fem   & 0.526 & 2.1    & 3.99       & 7    & 2.18 & 3.02 & 4.44 & 4.07 \\
PFlow\_742       & 0.743 & 18.941 & 25.5       & 58   & 6.16 & 12.5 & 11.4 & 17.7 \\
Serena           & 1.391 & 32.962 & 23.69      & 201  & 9.96 & 13.4 & 33.1 & 32.1 \\
StocF-1465       & 1.465 & 11.235 & 7.67       & 80   & 6.48 & 10.5 & 13.4 & 17.0 \\
thermal2         & 1.228 & 4.904  & 3.99       & 10   & 3.94 & 4.40 & 12.3 & 13.5 \\
tmt\_sym         & 0.727 & 2.904  & 4          & 5    & 2.45 & 2.98 & 4.54 & 4.97
\end{tabular}
\label{tab:matrices-times}
\end{table*}

We use CUDA 10.2 for the NVIDIA V100 experiments. The only exception is that we used CUDA 9.2 in the CUSP vs. Kokkos Kernels comparisons since CUSP does not officially support CUDA 10. For the AMD MI100 experiments, we used the HIP backend of Kokkos and the ROCM 4.3.0 toolchain. For Intel CPU results, we used a dual-socket Intel Xeon Platinum 8160 system with 24 cores per socket and 2 threads per core, and compiled with Intel 20.2. Finally, ARM CPU results were measured on a dual-socket Cavium ThunderX2 with 28 cores per socket and 2 threads per core. We used armclang 20.1 on the ARM system. The MIS-2 implementation used in these experiments is available in the Kokkos Kernels library \cite{kk}. Summary statistics for the 17 matrices and the average running times on each of these four platforms can be found in Table \ref{tab:matrices-times}. 48 and 56 OpenMP threads were used on the Skylake and ThunderX2 systems respectively, since these gave the lowest wall time (thread scalability is discussed in section \ref{sec:scalability}).

\subsection{Impacts of Algorithmic Optimizations}
\label{sec:incremental-opt}

\begin{figure}[t]
  \vspace{-0.5cm}
  \centerline{\includegraphics[width=0.50\textwidth]{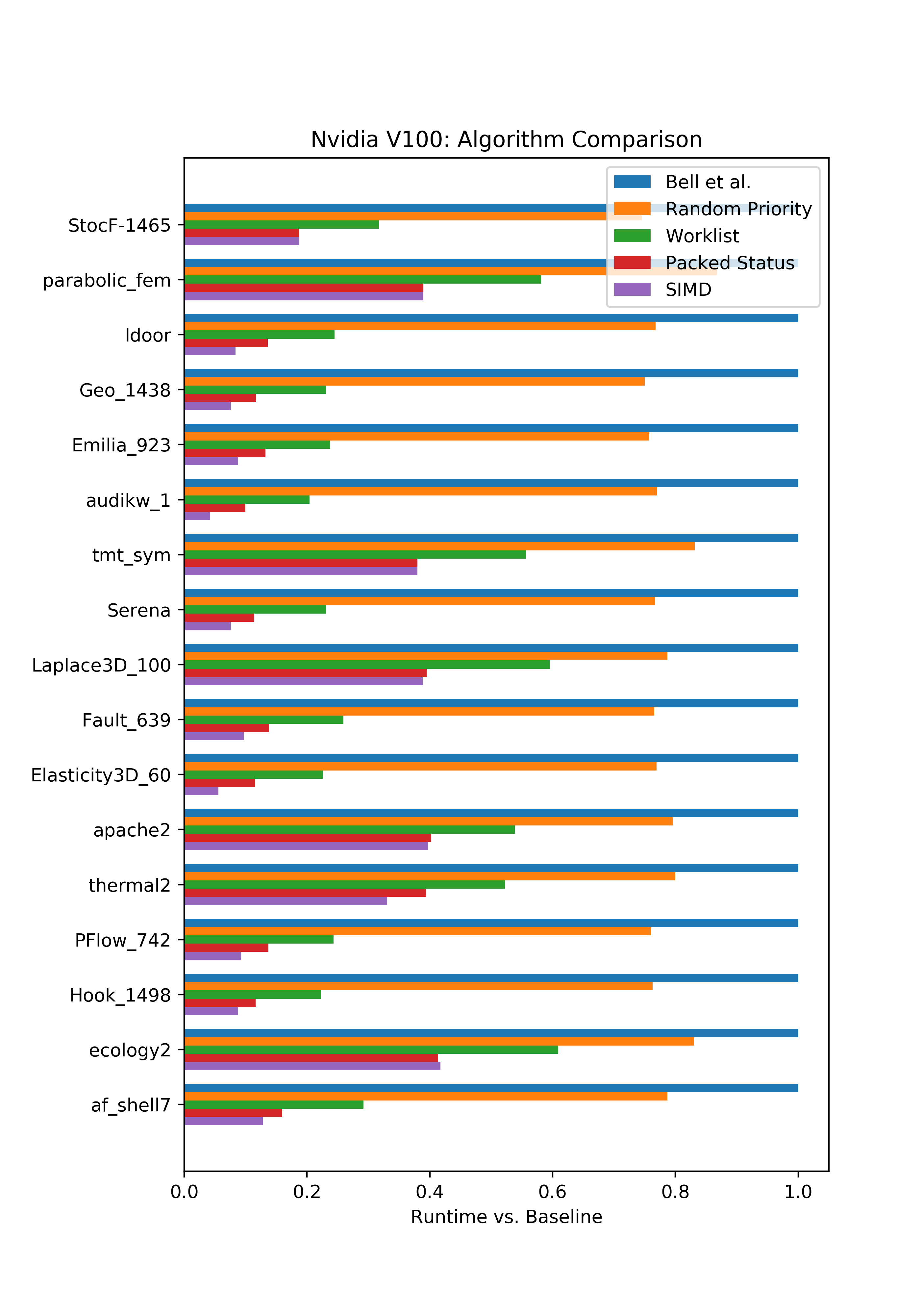}}
  \vspace{-0.6cm}
  \caption{Cumulative speedups from four optimizations over the Kokkos baseline implementation \cite{bell} for the 17 matrices on NVIDIA V100.}
  \label{fig:incremental-opt}
  \vspace{-0.5cm}
\end{figure}

Fig. \ref{fig:incremental-opt} shows the effects of the four optimizations discussed in Section \ref{sec:optimizations} on 17 matrices. We use NVIDIA V100 architecture for these comparisons. Five different implementations are compared. The first is a reference implementation of Bell's algorithm for a general MIS-k, which we call with $k=2$ \cite{bell}. Each of the next four implementations adds another optimization from section \ref{sec:optimizations}, while retaining all previous optimizations. The final ``SIMD'' implementation is Algorithm \ref{alg:kk} as implemented in Kokkos Kernels, with all four optimizations.

All optimizations provide significant speedups, but worklists provide the greatest speedup with 2.55x (geometric mean) on these 17 graphs. 
Random priority, Packed Status, and SIMD, lead to 1.28x, 1.72x, 1.37x speedups respectively.
\emph{The four optimizations combined yield a 8.97x speedup}. 
Notice that for graphs with $|E|/|V| < 16$  ``SIMD'' and ``Packed Status'' versions run in equal time, as the heuristic is disabling SIMD. 

\subsection{Algorithm Scaling on Structured Problems}

\begin{table}[h]
\caption{MIS-2 Size and iteration count for varying problem sizes using Structured Problems}
\centering
\begin{tabular}{l|lll}
Problem             & $|V|$       & $|$MIS-2$|$ & Iters  \\ \hline
Elasticity 30x30x30 & 81000 & 634   & 8     \\
Elasticity 60x30x30 & 162000 & 1291 & 10    \\
Elasticity 60x60x30 & 324000 & 2454  & 10    \\
Elasticity 60x60x60 & 648000 & 4833  & 10    \\
Laplace 50x50x50    & 125000 & 11469 & 9     \\
Laplace 100x50x50   & 250000 & 22909 & 9     \\
Laplace 100x100x50  & 500000 & 45333 & 9     \\
Laplace 100x100x100 & 1000000 & 90041 & 10
\end{tabular}
\label{tab:structured-scaling}
\end{table}

Table \ref{tab:structured-scaling} gives the MIS-2 size and number of iterations required for Algorithm \ref{alg:kk} for varying grid sizes of structured matrices from two different problem types generated with the Galeri package of Trilinos. We use structured problems here as it is easy to vary the grid dimension and see how the quality of the results and the number of iterations needed for MIS-2 vary in response. Notice that the average degree of the Elasticity and Laplace matrices are 81 and 7 respectively. This results in 0.7\% and 9\% of vertices in MIS-2. The results show that the number of iterations  goes up by 1-2 iterations as the grid size increases 4x-8x, and for a given problem type the MIS-2 size remains proportional to $|V|$. \emph{This demonstrates good algorithmic scaling as graph sizes increase}.

\subsection{Performance Portability and Scalability}
\label{sec:scalability}

\begin{figure}[h]
  \vspace{-0.3cm}
  \centering
  \centerline{\includegraphics[width=0.4\textwidth]{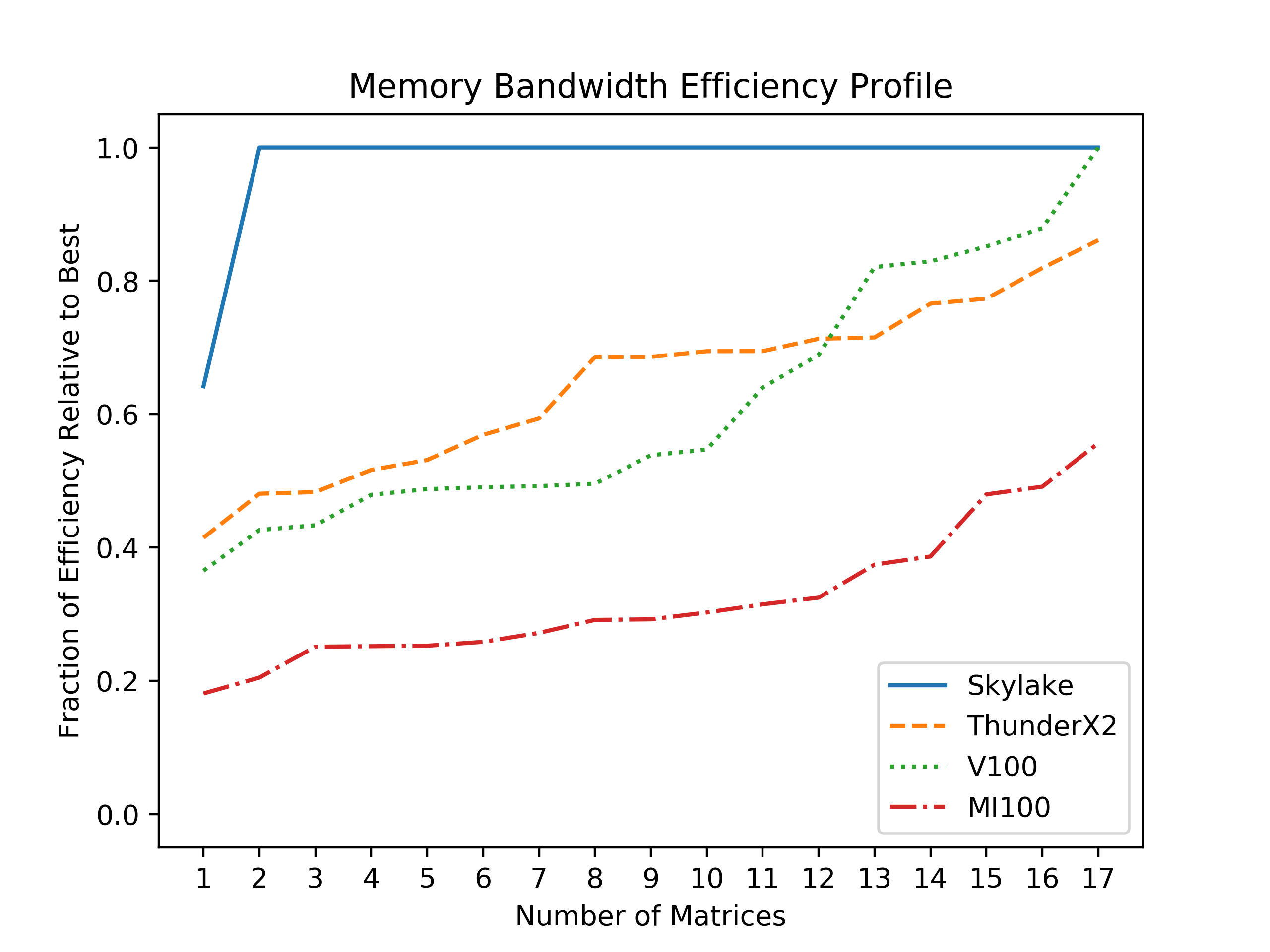}}
  \caption{Bandwidth efficiency profiles of all four architectures on the 17 problems. }
  \label{fig:portability}
  \vspace{-0.1cm}
\end{figure}

Fig. \ref{fig:portability} shows a profile of the portability of our MIS-2 implementation on the four architectures. Our algorithm is memory bound, so ideal portability would mean running times inversely proportional to each device's theoretical memory bandwidth. The global memory bandwidth is 1200 GB/s on the MI100, and 900 GB/s on the V100. The Intel Skylake system has 238 GB/s of main memory bandwidth and the ThunderX2 ARM system has 317 GB/s \cite{SystemBandwidth}. Here, we define the bandwidth efficiency as the number of MIS-2 instances computed per second, divided by the bandwidth. Given perfect portability, this number would be the same across all platforms. For each system and each problem, the \emph{fraction of efficiency} in the profile plot is the efficiency, divided by the best efficiency for that problem among the four platforms. In practice, the Intel Skylake has the highest efficiency on all but one problem.

\begin{figure}[t]
  \centering
  \centerline{\includegraphics[width=0.4\textwidth]{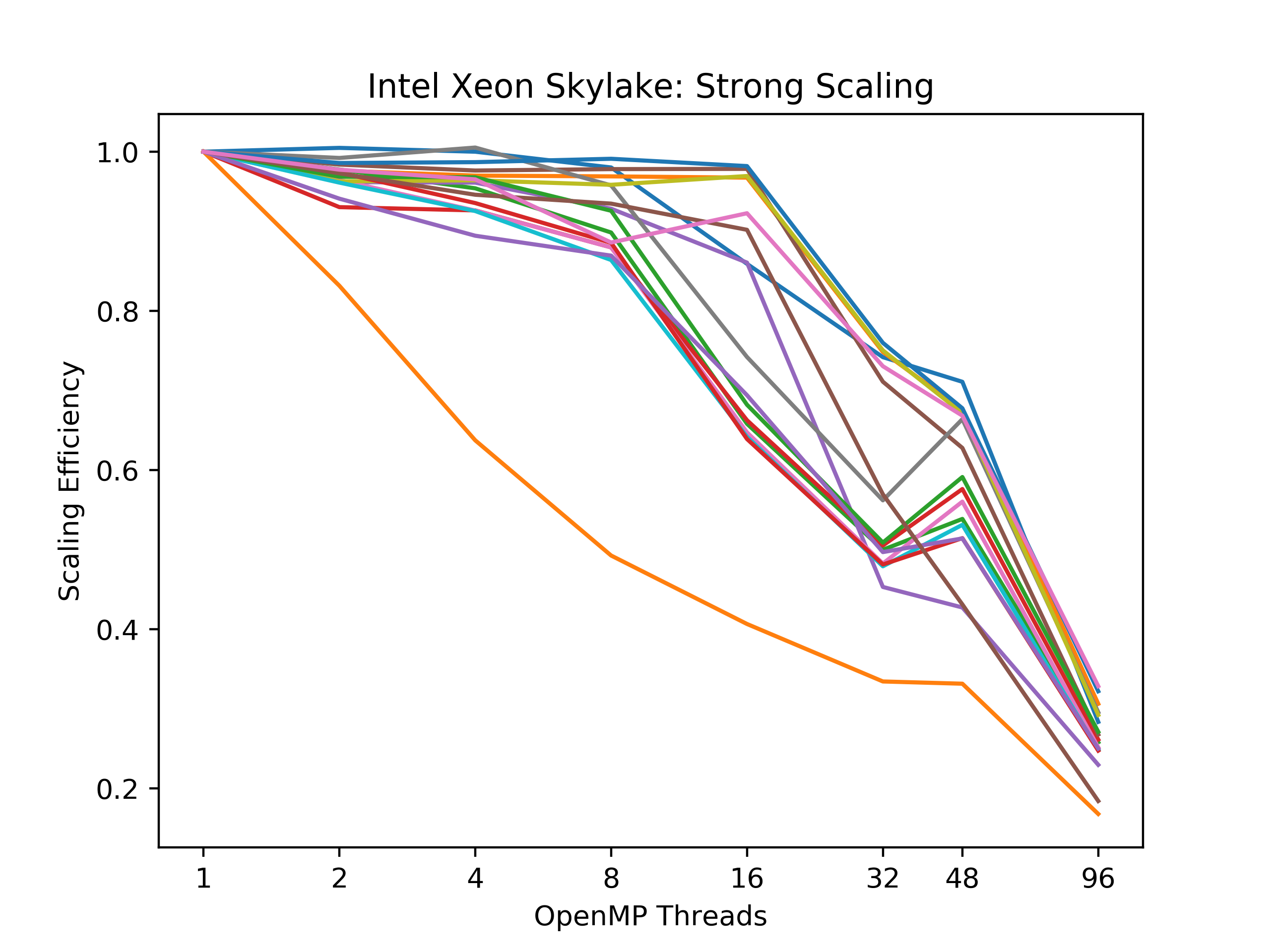}}
  \caption{Strong scaling efficiency of MIS-2 on dual Intel Xeon Platinum 8160 for the 17 graphs. Ideal scaling is 1, where $\mathit{time} \propto 1/\mathit{threads}$. Each additional thread up to 48 (number of total physical cores) gives a speedup, but if we attempt to use all 96 hyperthreads, the algorithm slows down.}
  \vspace{-0.2cm}
  \label{fig:blake}
\end{figure}

\begin{figure}[t]
  \vspace{-0.1cm}
  \centering
  \centerline{\includegraphics[width=0.4\textwidth]{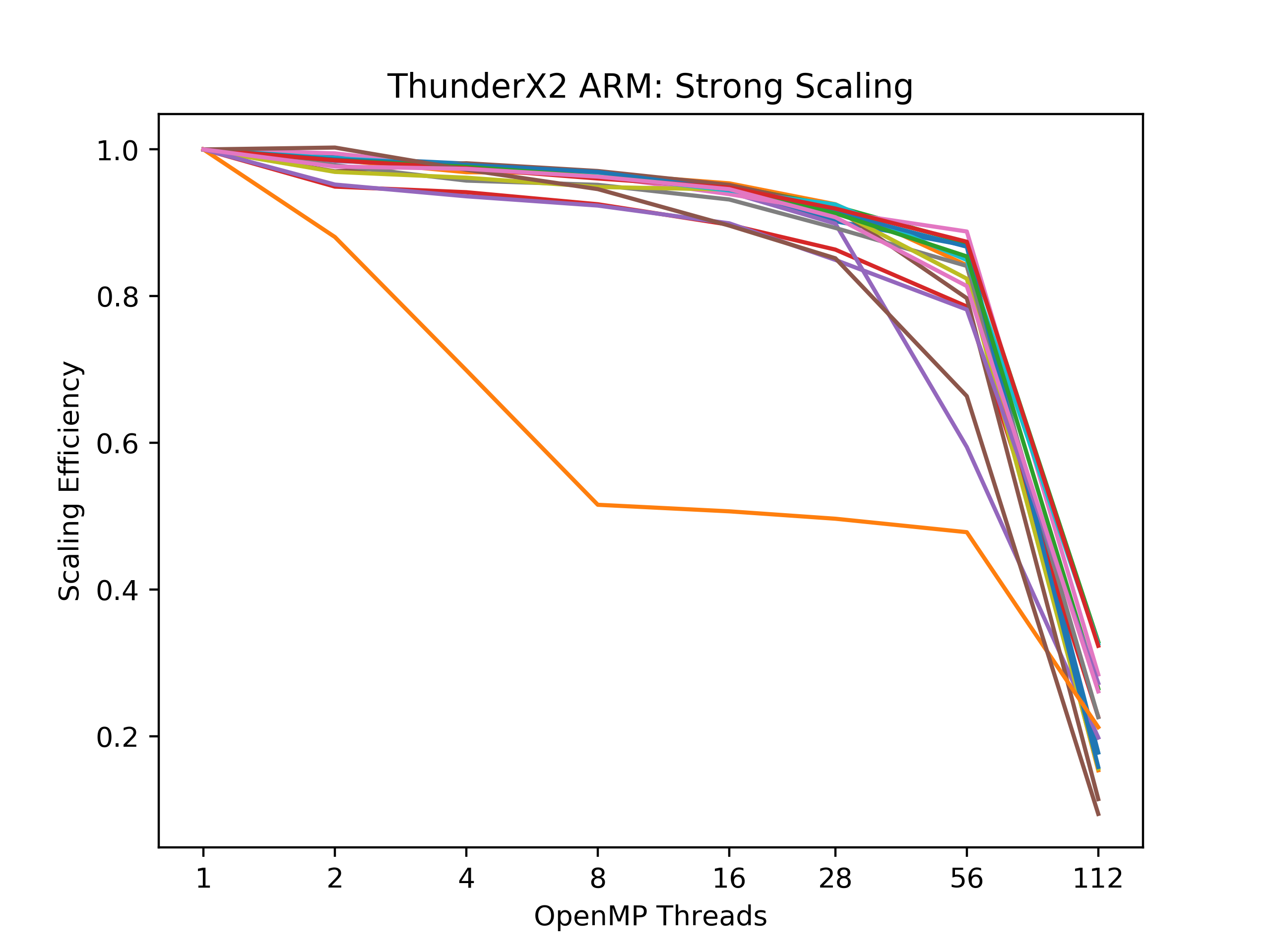}}
  \caption{Strong scaling efficiency of MIS-2 on dual ThunderX2 ARM CPU for the 17 graphs. Very good scaling is achieved up to the 56 physical \emph{cores}, but as with Intel, using all 112 hardware \emph{threads} does not yield a speedup.}
  \label{fig:stria}
  \vspace{-0.2cm}
\end{figure}

Figs. \ref{fig:blake} and \ref{fig:stria} demonstrate the strong scaling of the Kokkos Kernels MIS-2 on Intel and ARM CPUs, respectively. Each line corresponds to one of 17 matrices and relates the scaling efficiency to the number of OpenMP threads. The ideal efficiency is 1, where the running time is inversely proportional to the number of threads. Both systems have two sockets and two threads per physical core, our MIS-2 algorithm scales well to the total number of physical cores (48 for Intel and 56 for ARM). \emph{We observe  26.9x and 43.9x speedup (geometric mean) on 48 and 56 threads on Intel and ARM respectively.}

\subsection{Performance comparison with CUSP and ViennaCL}

\begin{figure}[h]
  \vspace{-0.25cm}
  \centering
  \centerline{\includegraphics[width=0.4\textwidth]{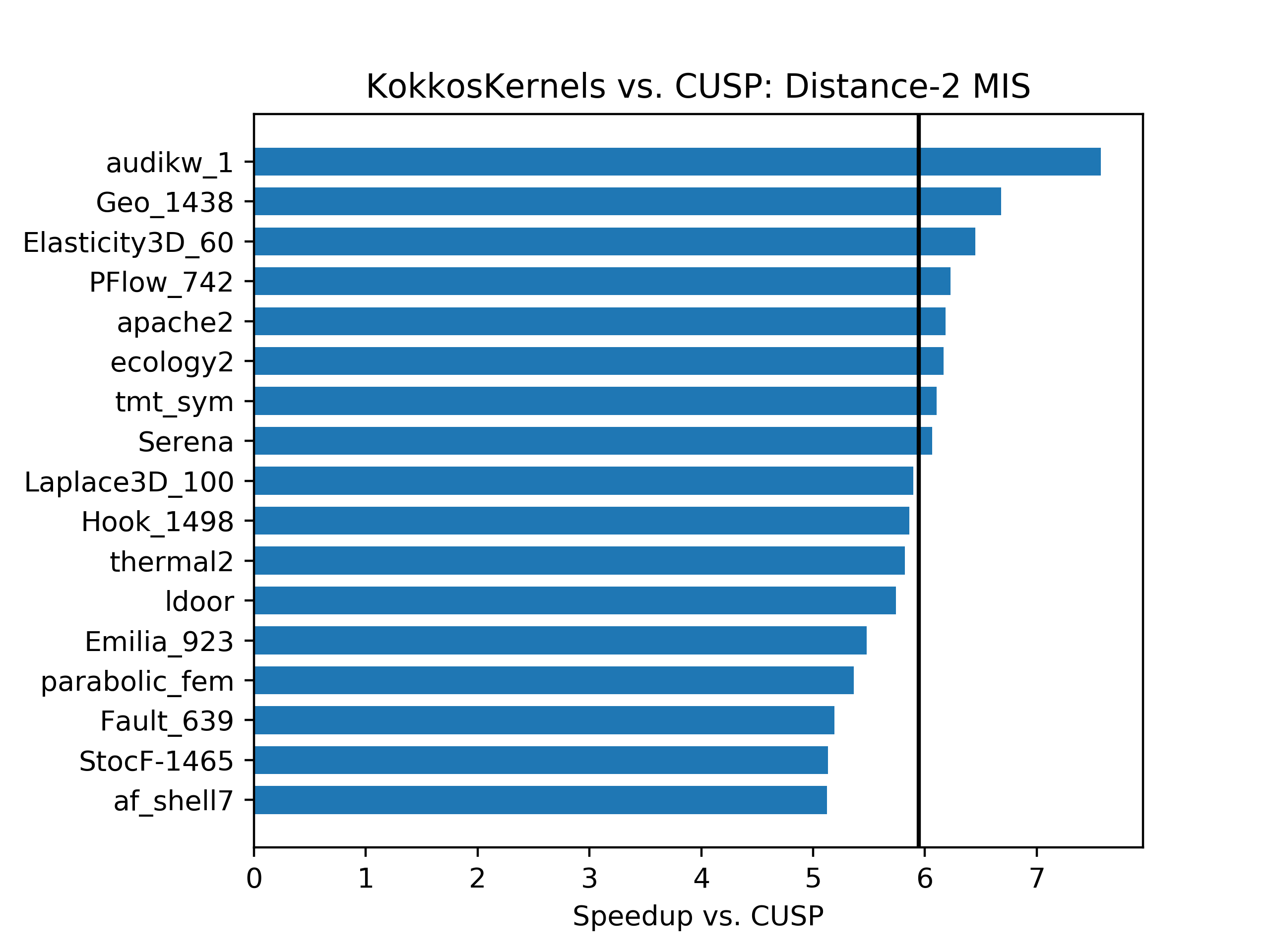}}
  \caption{MIS-2: Comparison of our approach with CUSP on V100 GPUs. The vertical line is the mean speedup.}
  \label{fig:cusp}
\end{figure}

\begin{figure}[h]
  \centering
  \vspace{-0.3cm}
  \centerline{\includegraphics[width=0.4\textwidth]{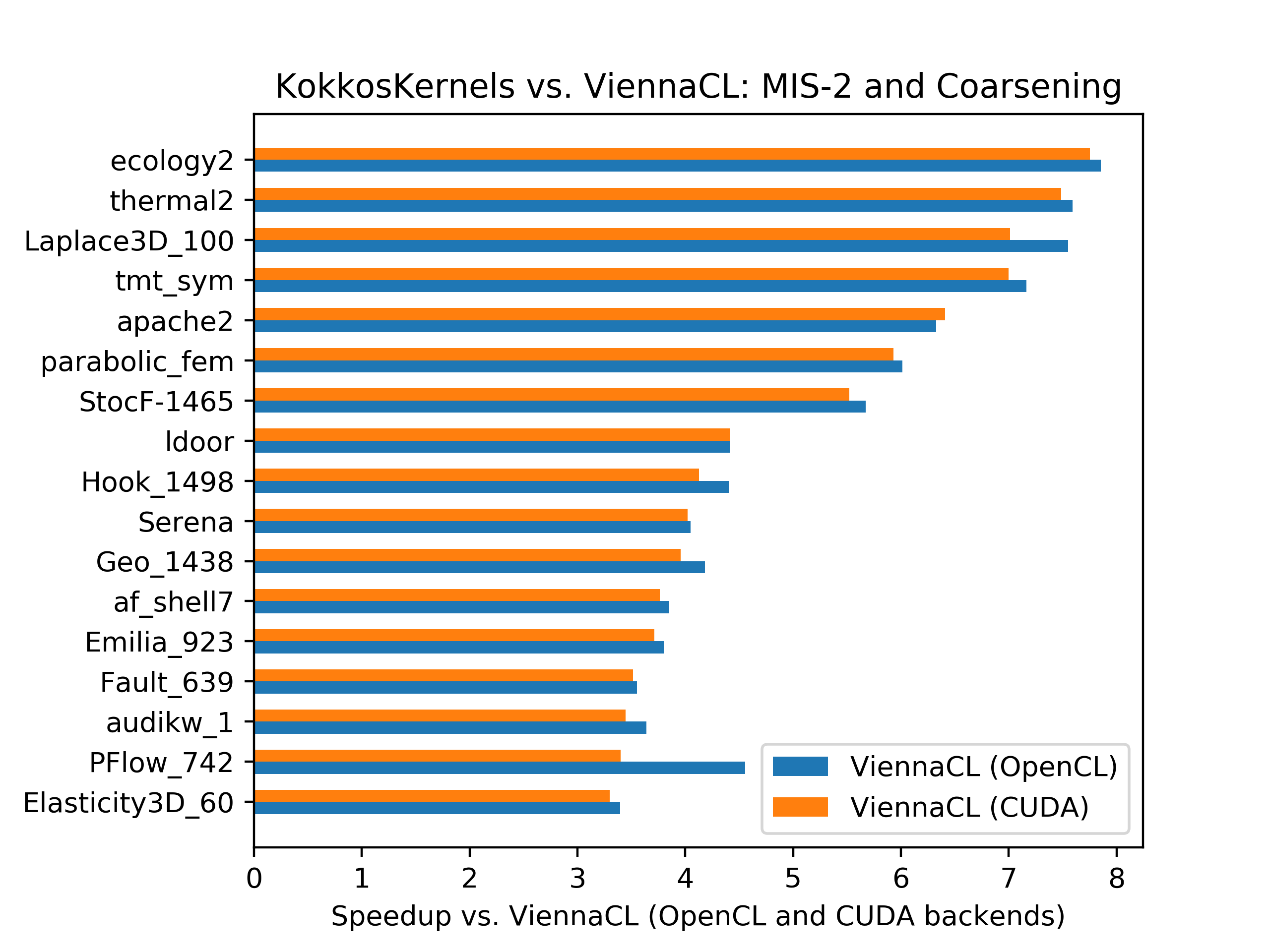}}
  \caption{MIS-2 based coarsening: Comparison of our approach with ViennaCL on V100 GPUs}
  \label{fig:vienna}
\end{figure}

Figs. \ref{fig:cusp} and \ref{fig:vienna} demonstrate our implemenation against two previous state-of-the-art implementations of Bell's algorithm on V100 GPUs. The first compares our implementation against CUSP \cite{cusp} for the problem of computing an MIS-2 alone. The second compares our MIS-2 plus the coarsening Algorithm \ref{alg:coarsen} against the ViennaCL library \cite{viennacl}. This is because ViennaCL exposes an interface for coarsening, but not for MIS-2 by itself. In both problems, and for both the CUDA and OpenCL backends of ViennaCL, \emph{our algorithm achieves 3-8x speedup on all seventeen matrices tested compared to ViennaCL. Kokkos Kernels MIS-2 achieves 5-7x speedup compared to CUSP on the 17 matrices.}

\subsection{Quality Comparison with CUSP and ViennaCL}
\begin{table}[h]
\caption{Quality of MIS-2: Number of vertices in MIS-2 for Kokkos Kernels, CUSP, and ViennaCL (higher is better)}
\centering
\begin{tabular}{l|lll}
                 & KK & CUSP   & ViennaCL \\ \hline
af\_shell7       & 9708 & 9742   & \textbf{9772}    \\
ecology2         & 139431     & \textbf{140110} & 139813  \\
Hook\_1498       & \textbf{21469}      & 20966  & 21077  \\
PFlow\_742       & \textbf{64880}      & 64763  & 64767   \\
thermal2         & 118217     & \textbf{118426} & 118327   \\
apache2          & 67750      & 67802  & \textbf{67884}   \\
Elasticity3D & \textbf{4833}       & 4791   & 4784     \\
Fault\_639       & \textbf{7901}        & 7835   & 7877     \\
Laplace3D   & 90041     & \textbf{90198}  & 90180   \\
Serena           & \textbf{16575}      & 16451  & 16439    \\
tmt\_sym         & 68827      & 68769  & \textbf{68835}  \\
audikw\_1        & \textbf{4263}       & 4201   & 4186    \\
Emilia\_923      & \textbf{11445}      & 11420  & 11427   \\
Geo\_1438        & 18168      & \textbf{18218}  & 18161   \\
ldoor            & \textbf{12464}      & 12326  & 12369   \\
parabolic\_fem   & 50396       & 50526  & \textbf{50530}   \\
StocF-1465       & \textbf{83419}       & 83401  & 83274
\end{tabular}
\label{tab:quality}
\vspace{-0.2cm}
\end{table}

Table \ref{tab:quality} compares the sizes of the MIS-2 produced by Kokkos Kernels algorithm (Algorithm \ref{alg:kk} / KK), CUSP, and ViennaCL. For all graphs, size of MIS-2 from all the three implementations are very similar. \emph{This shows our faster, portable, implementation achieves the same quality results as the other algorithms and their native implementations}.

\subsection{Multigrid integration and quality of aggregates}
\label{sec:aggregation-results}

\begin{table}[h]
\caption{Summary of MueLu Results for several aggregation algorithms on $100^3$ Laplace3D problem.}
\centering
\begin{tabular}{l|lllll}
                & Iters & Agg. & Setup & Solve & Det. \\ \hline
Serial Agg      & 25      & 0.673  & 2.80    & 0.390 & \checkmark  \\
Serial D2C & 23       & 0.125  & 0.601   & 0.383 & \\
NB D2C     & 31.3 & 0.274 & 0.734   & 0.447 &   \\
MIS2 Basic    & 49      & 0.0226 & 0.471   & 0.562 & \checkmark  \\
MIS2 Agg  & 22 & 0.0352  & 0.538   & 0.370  & \checkmark
\end{tabular}
\label{tab:muelu}
\vspace{-0.3cm}
\end{table}

In this subsection we compare our two MIS-2 based aggregation technique against three existing techniques in the MueLu multigrid package of Trilinos \cite{muelu}. The techniques are:

\begin{itemize}
    \item \textbf{Serial Agg}: MueLu's original aggregation algorithm. All parts of setup execute sequentially on the host CPU. Similar to ML's non-MIS2 aggregation \cite{ML}, but also includes enhancements designed by Wiesner \cite{TobiasAgg}.
    \item \textbf{Serial D2C}: Uses the Kokkos Kernels serial implementation of net-based distance-2 graph coloring \cite{D2Coloring}. The vertices of a given color form a distance-2 independent set. For each color, each vertex of that color is treated as a root and forms an aggregate with its neighbors. Like our Algorithm \ref{alg:agg}'s phase 2, a root only forms an aggregate if it has sufficiently many unaggregated neighbors. The coloring is reverse offloaded to host, but the aggregation is done in parallel. The way leftover vertices are joined to aggregates makes this algorithm nondeterministic.
    \item \textbf{NB D2C}: The same as Serial D2C, except the coloring is also computed on device using a parallel implementation of net-based coloring.
    \item \textbf{MIS2 Basic}: Algorithm \ref{alg:coarsen}.
    \item \textbf{MIS2 Agg}: Algorithm \ref{alg:agg}.
\end{itemize}


Each experiment sets up a multigrid V-cycle SA preconditioner using the specified aggregation algorithm to coarsen at all levels. The preconditioner was then used to solve a $100^3$ Laplace3D problem generated by MueLu to a tolerance of $10^{-12}$, using 2 sweeps of the Jacobi method as a smoother and conjugate gradient (CG) as the main solver. The experiments were run on a Power9 and NVIDIA V100 system. All quantities are averaged over 50 trials.

Table \ref{tab:muelu} presents the results. ``Iters'' is the number of iterations of the CG solver required to converge. ``Agg'', ``Setup'' and ``Solve'' are the total times spent in aggregation, MueLu setup, and the preconditioned solve respectively. ``Det.'' contains a checkmark for schemes which are deterministic.
The main reason for Serial Agg being much slower in setup is that it takes a non-Kokkos (i.e. sequential) path in MueLu, while  the four others do not. During the solve, all five schemes use the same Kokkos-based path. Our new algorithm MIS-2 Agg is 22x faster in aggregation than Serial Agg and results in fewer iterations. Compared to parallel distance-2 coloring (NB D2C) and MIS2 Basic, it results in 29\% and 55\% fewer iterations ad as a result reduce solve time.   Aside from MIS2 Basic, MIS2 Agg now has the fastest \emph{deterministic} setup by a factor of 5.2, and the fastest overall setup by 12\%.

\subsection{Cluster Multicolor Gauss-Seidel}
\label{sec:cmgs-results}

\begin{table}[t]
\caption{Point vs. Cluster Multicolor Gauss-Seidel as preconditioners for GMRES. We compare setup and total apply time, as well as GMRES iterations.}
\centering
\begin{tabular}{l|llll}
                 & P. Setup & C. Setup & P. Apply & C. Apply \\ 
                & &  & (P. Iters) & (C. Iters) \\ \hline
bodyy5           & 0.0154      & \textbf{0.00849}       & 0.124 (187.0)     & \textbf{0.0616} (\textbf{172.6}) \\
Elasticity3D\_60 & 0.174       & \textbf{0.0438}        & 7.41 (\textbf{328.2})      & \textbf{4.56} (337.4) \\
Geo\_1438        & 0.209       & \textbf{0.0662}        & 11.1 (408.5)     & \textbf{4.73} (\textbf{388.4}) \\
Laplace3D\_100   & 0.0553      & \textbf{0.0409}        & 0.664 (158.4)    & \textbf{0.567} (\textbf{144.6}) \\
Serena           & 0.215        & \textbf{0.0664}        & 6.55 (227.0)      & \textbf{2.93} (\textbf{219.2})

\end{tabular}
\vspace{-0.2cm}
\label{tab:cgs}
\end{table}

In Table \ref{tab:cgs}, we compare the Kokkos Kernels implementation of point multicolor symmetric Gauss-Seidel (SGS) \cite{MehmetColoring} against our cluster multicolor SGS method based on Algorithm \ref{alg:cgs}, with Algorithm \ref{alg:agg} used for coarsening. The SGS methods are used as preconditioners for a GMRES solver. Only systems which converge to a tolerance of $10^-8$ within 800 iterations are included. Times were measured on a V100 GPU and averaged over 50 trials. In all five systems, the cluster method results in a speedup for both the setup and apply. For both methods, the setup time is dominated by greedy graph coloring \cite{MehmetColoring}, but the cluster method colors a smaller, coarsened graph. The apply speedup for cluster is partly explained by the cluster method reducing the number of iterations by 5\% (geometric mean).

\section{Conclusion and Future Work}
We presented a portable, parallel, deterministic algorithm for the distance-2 maximal independent set problem. We also demonstrated that our implementation is faster than the existing state of the art implementations, and is performance portable across several CPU and GPU architectures. We isolated four key optimizations that explain the speedup, and established more generally how they can be applied to a wide variety of parallel graph algorithms. We developed a graph coarsening scheme using the parallel MIS-2 algorithm. We showed that  our aggregation lowers setup time even compared to other non-deterministic schemes in a multigrid preconditioner \cite{muelu}. We also describe an extension to the multicolor Gauss-Seidel parallel preconditioner which uses our MIS-2 based coarsening to improve convergence and reduce both setup and apply time.
In the future, we would like to see our aggregation scheme integrated into the algebraic multigrid in the PETSc library, since it already uses many other features in Kokkos Kernels \cite{petsc}. We also plan to evaluate our graph coarsening algorithm in the context of multilevel graph partitioning as a replacement for the MIS-2 based coarsening algorithm of Bell et al. \cite{bell} as used in Gilbert et al. \cite{Gilbert} in the future.

\section{Acknowledgment}
We thank Mike Gilbert (Pennsylvania State University) for creating the initial reference implementation of the Bell/Dalton/Olson MIS-2 algorithm.
Funding was provided by the Exascale Computing Project.
Sandia National Laboratories is a multimission laboratory managed and operated by National Technology and Engineering Solutions of Sandia, LLC, a wholly owned subsidiary of Honeywell International, Inc., for the U.S. Department of Energy's National Nuclear Security Administration under contract DE-NA-0003525.

\bibliographystyle{plain}
\bibliography{main}

\end{document}